\documentstyle[epsfig,12pt,preprint,tighten,aps]{revtex}
\begin{document}

\title{\rightline{{\tt 1st of Dec 1999}}
\ \\
On the sign of the neutrino asymmetry induced by 
active-sterile neutrino oscillations in the early Universe}
\author{P. Di Bari$^{1}$ and R. Foot$^2$}
\date{1st of December}
\maketitle
\begin{center}
{\em
$^1$ Dipartimento di Fisica, Universit\`a di Roma ``La Sapienza" and\\
I.N.F.N Sezione di Roma 1\\
P.le Aldo Moro, 2, I00185 Roma, Italy\\
(dibari@roma1.infn.it)\\
$^2$ School of Physics \\
Research Centre for High Energy Physics\\
The University of Melbourne\\
Parkville 3052 Australia\\
(foot@physics.unimelb.edu.au)}
\end{center}
\vspace{-1cm}
\begin{abstract}
We deal with the problem of the final sign of the neutrino 
asymmetry generated
by active-sterile neutrino oscillations in the Early Universe
solving the full momentum dependent quantum kinetic equations. 
We study the parameter region $10^{-2} \stackrel{<}{\sim} 
|\delta m^2|/eV^2\le 10^3$.
For a large range of $\sin^2 2\theta_0$ values
the sign of the neutrino asymmetry 
is fixed and does not oscillate.
For values of mixing parameters in the region
$10^{-6}\stackrel{<}{\sim}\sin^{2}2\theta_{0}\stackrel{<}{\sim}
3\times 10^{-4}\,({\rm eV}^{2}/|\delta m^{2}|)$, 
the neutrino asymmetry appears to undergo rapid oscillations
during the period where the exponential growth occurs. 
Our numerical results indicate that the oscillations are
able to change the neutrino asymmetry sign. The sensitivity
of the solutions and in particular of the final sign of 
lepton number to small changes in the initial conditions 
depends whether the number of oscillations is high enough.
It is however not possible to conclude whether this effect is
induced by the presence of a numerical error or is an intrinsic 
feature. As the amplitude of the statistical fluctuations 
is much lower than the numerical error, our numerical analysis
cannot demonstrate the possibility of a chaotical generation of
lepton domains. In any case this possibility is confined
to a special region in the space of mixing parameters
and it cannot spoil the compatibility of the
$\nu_{\mu}\leftrightarrow\nu_{s}$ solution to the neutrino 
atmospheric data obtained assuming a small mixing of the $\nu_{s}$
with an ${\rm eV}-\tau$ neutrino.
\end{abstract}

\newpage

\section{Introduction}

If light sterile neutrinos exist, then this will lead
to important implications for early Universe cosmology.
This is because ordinary-sterile neutrino oscillations
generate large neutrino asymmetries for the large range
of parameters, $\delta m^2 \stackrel{>}{\sim} 10^{-5}\
eV^2, \ \sin^2 2\theta_0 \stackrel{>}{\sim} 
10^{-10}$\cite{ftv,fv1,fv2,f,bvw,pas,nfb}. This is a generic feature 
of ordinary-sterile neutrino oscillations for this 
parameter range. For $\delta m^2 \stackrel{<}{\sim}
10^{-5}\ eV^2$, the evolution of neutrino asymmetry
is qualitatively quite different as collisions 
are so infrequent and a large neutrino asymmetry
cannot be generated\cite{barb,kir}.
Interestingly, some people do not currently accept that
large neutrino asymmetry
is generated in the early Universe\cite{dolgov}.
We will comment briefly on this later in the paper.

An important issue which has yet to be fully addressed
is the sign of this asymmetry. Is it always fixed or
can it be random? 
This is an important issue because a random
asymmetry may lead to domains with lepton number of 
different signs\cite{ftv}. If such domains exist then
this may lead to observable consequences. For example,
active neutrinos crossing 
the boundaries of these lepton domains could
undergo a MSW resonance which would
lead to a new avenue of sterile neutrino production\cite{shifu}.
In Refs.\cite{shi,eks,Sorri99} the issue of sign of the asymmetry
was discussed in the approximation that all of the neutrinos 
have the same momentum (i.e. $p = 3.15T$ instead of the
Fermi-Dirac distribution).
This approximation is not suitable for discussing
the temperature region where the exponential growth in
neutrino asymmetry occurs. 
The reason is that in the average momentum toy-model,
all of the neutrinos enter the MSW resonance 
at the same time which significantly enhances
the rate at which neutrino asymmetry is created at
$T = T_c$.
The rapid creation of neutrino asymmetry significantly reduces
the region where the oscillations are adiabatic\cite{fv1}. 

Thus it is clear that the neutrino momentum
dependence must be properly taken into account.
This was done in Ref.\cite{fv1}
where an approximate solution to the quantum kinetic
equations was derived. This approximate
solution was called the `static approximation' and
was re-derived in a different way in Ref.\cite{bvw} where
it was shown that this approximation was just
the adiabatic limit of the quantum kinetic
equations (QKE's) in the region
where lepton number generation is dominated by collisions.
Anyway, in the limit where this approximation is valid,
it was shown in Ref.\cite{fv1} that the sign is 
completely fixed. 
The static approximation is valid for a large range of parameters
but is not valid for large $\sin^2 2\theta_{0} \stackrel{>}{\sim}
10^{-6}, \ \delta m^2 \sim -10\ eV^2$. It breaks down in 
this region because the neutrino asymmetry is generated
so rapidly during the exponential growth phase that the quantum kinetic
equations are no longer adiabatic.
Thus, while Ref.\cite{fv1} partially answers the question of
sign, it does not give the complete answer. 
The purpose of this paper is to examine the issue 
of the sign of the asymmetry by numerically solving
the quantum kinetic equations. 

The outline of this paper is as follows:
In section 2 we present some necessary preliminary
discussion on active-sterile neutrino oscillations
in the early Universe.
In section 3 we examine the likely size of the statistical
fluctuations in the early Universe. In section 4
we describe the numerical results of our study of 
the region of parameter space where the sign of the neutrino
asymmetry is fixed.  Using the results 
of section 3, we are able to conclude that 
in this region the statistical
fluctuations cannot have any effect and 
the generated lepton number would have the same 
sign in all the points of space. 
In section 5 we describe the features of the transition
from the region with no oscillations to one 
where the neutrino asymmetry oscillates 
for a short period during the exponential growth.
In section 6 we conclude.
Also included is an appendix giving some numerical
details, which we hope will be
useful to other workers in the field such as 
the authors of Ref.\cite{dolgov}.
 
\section{Preliminary discussion}

Our notation/convention for ordinary-sterile 
neutrino two state mixing is as follows. The weak
eigenstates $\nu_{\alpha}$ ($\alpha = e, \mu$ or $\tau$) 
and $\nu_{s}$ are linear combinations
of two mass eigenstates $\nu_a$ and $\nu_b$,
\begin{equation}
\nu_{\alpha} = \cos\theta_0 \nu_a + 
\sin\theta_0 \nu_b,\
\nu_{s} = - \sin\theta_0 \nu_a + 
\cos\theta_0 \nu_b,
\label{zwig}
\end{equation}
where $\theta_0$ is the vacuum mixing angle.
We define $\theta_0$ so that 
$\cos2 \theta_0 > 0$ and
we adopt the convention that $\delta m^2_{\alpha\beta'} 
\equiv m^2_b - m^2_a$.
Recall that the $\alpha$-type neutrino asymmetry is defined by 
\begin{equation}
L_{\nu_\alpha} \equiv 
{n_{\nu_\alpha} - n_{\overline \nu_\alpha} \over n_\gamma}.
\label{def}
\end{equation}
In the above equation, $n_{\gamma}$ is the number density of photons.
Note that when we refer to ``neutrinos'', sometimes we
will mean neutrinos and/or antineutrinos.
We hope the correct meaning will be clear from context.
Also, if neutrinos are Majorana particles, then technically
they are their own antiparticle. 
Thus, when we refer to ``antineutrinos'' we obviously
mean the right-handed helicity state in this case.

The density matrix\cite{stod,bm} for an ordinary 
neutrino, $\nu_\alpha$ ($\alpha =
e, \mu, \tau$), of momentum $p$ oscillating with a sterile neutrino
in the early Universe can be parameterized as follows:
\begin{equation}
\rho_{\alpha\beta'} (p) = {1 \over 2} [P_0(p)I + {\bf P}(p)\cdot
{\bf \sigma}],\quad
\overline{\rho}_{\alpha\beta'}(p) = 
{1 \over 2} [\overline{P}_0(p)I + {\bf \overline{P}}(p)\cdot
{\bf \sigma}],
\label{kdf}
\end{equation}
where $I$ is the $2 \times 2$ identity matrix, the ``polarisation vector''
${\bf P}(p) = P_x(p){\bf \hat x} + P_y (p) {\bf \hat y}
+ P_z(p){\bf \hat z}$ and ${\bf \sigma} = \sigma_x {\bf \hat x} + 
\sigma_y {\bf \hat y} + \sigma_z {\bf \hat z}$, with 
$\sigma_i$ being the Pauli matrices.
It will be understood that the density matrices
and the quantities $P_i(p)$ also depend on time $t$ or, equivalently, 
temperature $T$. The time-temperature relation
for $m_e \stackrel{<}{\sim} T \stackrel{<}{\sim} m_\mu$ is
$dt/dT \simeq -M_P/5.44T^3$, where $M_P \simeq 1.22 \times 10^{22} \
{\rm MeV}$ is the Planck mass.  

We will normalise the density matrices so that
the momentum distributions of $\nu_{\alpha}(p)$ 
and $\nu_{s}(p)$ are given by
\begin{equation}
f_{\nu_{\alpha}}(p) = {1 \over 2}[P_0(p) + P_z(p)]f_{0}(p), \
\quad f_{\nu_{s}}(p) = {1 \over 2}[P_0(p) - P_z(p)]f_{0}(p),
\label{c}
\end{equation}
where
\begin{equation}
f_{0}(p) \equiv {1 \over 1 + \exp\left({p \over T}\right) },
\end{equation}
is the Fermi-Dirac distribution (with zero chemical potential).
Similar expressions pertain to antineutrinos (with
${\bf P}(p) \to {\bf \bar P}(p)$ and $P_0 \to \bar P_0$). 
The evolution of ${\bf P}(p), P_0(p)$ (or
${\bf \bar P}(p), \bar P_0(p)$) are governed by the 
equations \cite{stod,bm,stod2,bvw}
\begin{eqnarray}\label{eq:b1}
{d {\bf P} \over dt} & = &
{\bf V}(x) \times {\bf P}(x) - D(x)[P_x (x) {\bf \hat x} 
+ P_y (x){\bf \hat y}]+ {d P_0 \over d t}\, {\bf \hat z},\nonumber \\
{d P_0 \over d t} & \simeq &  
\Gamma (x)\left[{f^{eq}(x) \over f_0(x)} - {1\over 2}(P_0 (x) + P_z (x))
\right],
\label{yd}
\end{eqnarray}
where $D(x) = \Gamma (x)/2$ and $\Gamma (x)$ is
the total collision rate of the weak 
eigenstate neutrino of adimensional momentum $x\equiv p/T$ with the background 
plasma\footnote{ From Ref.\cite{ekt,bvw} it is given by $\Gamma (x) = 
yG_F^2T^5x$ where $y \simeq 1.27$ for $\nu = \nu_e$ and 
$y \simeq 0.92$ for $\nu = \nu_{\mu},\nu_{\tau}$ (for $m_e 
\stackrel{<}{\sim} T \stackrel{<}{\sim} m_\mu$).} 
and $f_{eq}(x)$ is the Fermi-Dirac distribution:  
\begin{equation}
f_{eq}(x) \equiv {1 \over 1 + \exp(x- \tilde{\mu}_{\alpha})}.
\end{equation}
where $\stackrel{\sim}{\mu_\alpha} \equiv \mu_{\nu_\alpha}/T$.
For anti-neutrinos,
$\stackrel{\sim}{\mu_{\alpha}} \to \stackrel{\sim}{\mu_{\bar
\alpha}} \equiv
\mu_{\bar \nu_{\alpha}}/T$.
The chemical potentials $\mu_{\nu_\alpha},
\ \mu_{\bar \nu_{\alpha}}$, depend on the neutrino asymmetry.  
In general, for a distribution in thermal equilibrium
\begin{equation}
L_{\nu_{\alpha}} = {1 \over 4\zeta (3)}\int^{\infty}_0
{x^2 dx \over 1 + e^{x-\stackrel{\sim}{\mu_{\alpha}}}} - 
{1 \over 4\zeta (3)}\int^{\infty}_0
{x^2 dx \over 1 + e^{x-\stackrel{\sim}{\mu_{\bar \alpha}}}},  
\end{equation}
where $\zeta (3) \simeq 1.202$ is the Riemann zeta function
of 3.  Expanding out the above equation,
\begin{equation}\label{j1}
L_{\nu_{\alpha}} \simeq {1 \over 24\zeta (3)}\left[
\pi^2 (\stackrel{\sim}{\mu}_{\alpha} - 
\stackrel{\sim}{\mu}_{\bar \alpha})
+6(\stackrel{\sim}{\mu}_{\alpha}^2 - 
\stackrel{\sim}{\mu}_{\bar \alpha}^2)\ln2
+ (\stackrel{\sim}{\mu}_{\alpha}^3 - 
\stackrel{\sim}{\mu}_{\bar \alpha}^3) \right],
\end{equation}
which is an exact equation for $\stackrel{\sim}{\mu}_{\alpha} =
- \stackrel{\sim}{\mu}_{\bar \alpha}$, otherwise it holds to a good
approximation provided that $\stackrel{\sim}{\mu}_{\alpha,\bar \alpha}
\stackrel{<}{\sim} 1$. For $T \stackrel{>}{\sim} T^{\alpha}_{dec}$
(where $T^e_{dec} \approx 2.5$ MeV and $T^{\mu,\tau}_{dec} \approx 3.5
$ MeV are the chemical decoupling temperatures),
$\mu_{\nu_{\alpha}} \simeq - \mu_{\bar \nu_{\alpha}}$
because processes such as $\nu_{\alpha} + \bar \nu_{\alpha}
\leftrightarrow e^+ + e^-$ are rapid enough to make
$\stackrel{\sim}{\mu}_{\alpha} +\stackrel{\sim}{\mu}_{\bar \alpha} 
\ \simeq \ \stackrel{\sim}{\mu}_{e^+} + \stackrel{\sim}{\mu}_{e^-} 
\simeq 0$.
However, for $1 MeV \stackrel{<}{\sim} T \stackrel{<}{\sim}
T^{\alpha}_{dec}$, weak interactions are rapid enough to 
approximately thermalise the neutrino momentum distributions,
but not rapid enough to keep the neutrinos in chemical equilibrium
\footnote{The chemical and thermal decoupling temperatures are so
different because the inelastic collision rates are much
less than the elastic collision rates. See e.g. Ref.\cite{ekt} for
a list of the collision rates.}.
In this case, the value of $\stackrel{\sim}{\mu}_{\alpha}$ is
approximately frozen at $T \simeq T^{\alpha}_{dec}$ (taking
for definiteness $L_{\nu_\alpha} > 0$), while the (negative) 
anti-neutrino chemical potential $\stackrel{\sim}{\mu}_{\bar \alpha}$
continues decreasing until $T \simeq 1$ MeV.


The quantity ${\bf V}(x)$ is given by\cite{bm,stod2}
\begin{equation} 
{\bf V}(x) = \beta (x) {\bf \hat x} + \lambda (x) {\bf \hat z},
\end{equation}
where $\beta (x)$ and $\lambda (x)$ are
\begin{equation}
\beta (x) = {\delta m^2 \over 2xT}\sin 2\theta_0,
\ \lambda (x) = -{\delta m^2 \over 2xT}[\cos2\theta_0 - b(x)\pm a(x)],
\label{sf}
\end{equation}
in which the $+(-)$ sign corresponds to neutrino (anti-neutrino)
oscillations. The dimensionless variables $a(x)$ and $b(x)$ 
contain the matter effects\cite{wolf78} (more precisely they are the
matter potential divided by $\delta m^2/2xT$).
For $\nu_{\alpha} \to \nu_s$ oscillations 
$a(x), b(x)$ are given by\cite{rn,ekm} 
\begin{equation}
a(x) \equiv {-4\zeta(3)\sqrt{2}G_FT^4L^{(\alpha)}x
\over \pi^2\delta m^2},
\ b(x) \equiv {-4\zeta (3) \sqrt{2} G_F T^6 A_{\alpha} x^2 \over
\pi^2 \delta m^2 M_W^2},
\label{sal}
\end{equation}
where $G_F$ is the Fermi constant, $M_W$ is the 
$W-$boson mass, $A_e \simeq 17$ and $A_{\mu, \tau} \simeq 4.9$ 
(for $m_e \stackrel{<}{\sim} T \stackrel{<}{\sim} m_\mu$).
The important quantity $L^{(\alpha)}$ is 
given by the following expression:
\begin{equation}
L^{(\alpha)} \equiv L_{\nu_\alpha} + L_{\nu_e} + L_{\nu_\mu}
+ L_{\nu_\tau} + \eta \equiv 2 L_{\nu_\alpha}+\tilde{L},
\end{equation}
where $\eta$ is a small term due to the asymmetry of the 
electrons and nucleons and is expected
to be very small, $|\eta| \sim 5\times 10^{-10}$. 
We will refer to $L^{(\alpha)}$ as the
{\em effective total lepton number}
(for the $\alpha$-neutrino species) since it really needs a name.
Note that the quantity $\tilde{L}$ is independent of
$L_{\nu_\alpha}$ 
and its value is currently unknown, but could presumably be
calculated within a model of baryo-leptogenesis\footnote{Alternatively,
it maybe set by divine intervention.}.
For our numerical work we took two different values
$\tilde{L} = 5\times 10^{-10}$ and $\tilde{L} = 5\times 10^{-11}$; in
this way we could verify that the final value of the total lepton number 
does not depend on this particular choice. 
The neutrino asymmetry $L_{\nu_\alpha}$, on the other hand,
evolves dynamically and is a function of ${\bf P}, P_{0}$ and
${\bf\bar{P}}, \bar{P}_{0}$.
The equations  Eq.(\ref{eq:b1}) constitute 
a closed set of differential equations for the 
8 distributions ${\bf P}, P_{0}, {\bf\bar{P}}, \bar{P}_{0}$. 
The `initial' conditions (for $T \to \infty$) are
simply $P_x = P_y = 0,\ P_0 = P_z = 1$ , assuming 
here for simplicity that 
initially $L_{\nu_\alpha} = 0$
(and similarly for antiparticles).
Note that the vanishing of $P_{x,y}$ is just
an example of the quantum Zeno effect.

From a computational point of view it is useful to 
write down an explicit differential 
equation for the neutrino asymmetry to be solved 
together (see also \cite{f}). This can be done 
considering that $dL_{\nu_\alpha}/dt=-dL_{\nu_s}/dt$ and, 
using the Eq.(\ref{c}), one easily gets:
\begin{equation}
{dL_{\nu_\alpha} \over dt} =
{1 \over 8\zeta (3)}\int
\beta(x) [P_y (x) - \bar P_y(x)] f_0(x)\,x^{2}dx.
\label{wed}
\end{equation}
We mainly employ the useful time saving approximation 
of integrating the oscillation equations, Eq.(\ref{eq:b1})
[and obviously also Eq.(\ref{wed})]
in the region around the MSW resonances. A detailed
description of the numerical procedures is given 
in the Appendix. Actually away from the resonance the 
oscillations are typically suppressed by the matter 
effects or by $\sin^2 2\theta_0$ (or both).  Thus, 
this should be a good approximation, which we 
carefully checked by taking larger slices of 
momentum space around the resonance 
(an example of this check is given in the Appendix).

Before we finish this section we would like
to briefly comment on the static approximation\cite{fv1,bvw}.
The static approximation can be derived by
solving the Quantum kinetic equations in the
adiabatic limit and assuming that the evolution
is dominated by collisions. The resulting 
equations are equivalent to 
keeping the differential equations
for $P_0, P_z, \bar P_0, \bar P_z$ [contained
in Eq.(\ref{yd})] which are functions of $P_y, \bar P_y$ 
which are given by
\begin{equation}
P_y = {-P_z \beta D \over D^2 + \beta^2 + \lambda^2}, 
\end{equation}
and $\bar P_y$ is similar to the above
equation except for the obvious replacement of
$P_z \to \bar P_z$ and $a \to -a$ in the definition of $\lambda$.
The evolution of $L_{\nu_\alpha}$ is obtained
from Eq.(\ref{wed}) above.
Numerically, the static approximation is much
easier to solve, but it is not always a valid
approximation. It is not valid at low temperatures
where coherent MSW transitions are important.
Also, if the neutrino asymmetry is created fast 
enough during the exponential growth phase then
it may not always be valid (which
is the case for $\delta m^2 \sim -10 \ eV^2$
and $\sin^2 2\theta_0 \stackrel{>}{\sim} 10^{-6}$, for 
example). 
In {\bf figures 1-3}, we give some numerical examples.
For {\bf fig. 1a}
we take $\nu_\tau \leftrightarrow \nu_s$ oscillations for
the parameter choices $\delta m^2 = -10 \ eV^2$,
$\sin^2 2\theta_0 = 10^{-7}$. Shown is the evolution
of the effective total lepton number, $L^{(\tau)}$,
with the solid line representing the numerical solution
of the quantum kinetic equations integrating
around the resonances
\footnote{Note that at low temperatures
$T \stackrel{<}{\sim} 5 \ MeV$ 
the repopulation must be
taken into account over the entire momentum range,
$0.01 < p/T < 20$.}. The dashed line
is the static approximation discussed above integrating
also around the resonances.
Observe that there is exact agreement between
the QKEs and the static approximation
when both are integrated around the resonances,
except at low temperatures 
where the static approximation
is not valid, as the evolution there
is dominated by coherent oscillations (MSW effect)
as discussed in detail in Ref.\cite{fv2}.
In {\bf fig.1b} we compare the static approximation
integrating around the resonance with the static approximation
integrating over the entire momentum range (which we
approximate to $0.01 < p/T < 20$).
There is some differences between the static
approximation integrated over the entire
momentum range with the static approximation (or QKE's) integrated
around the resonances.
The difference at high temperature can be qualitatively
understood because the resonances at high 
temperature occur at very low momentum, $p_{res}/T \stackrel{<}{\sim}
1$ where there are very few neutrinos.
Thus, the region around $p_{res}/T \sim 2$ can be
somewhat important despite being away from the resonance
because of the larger number of neutrinos.
Also note that the point where the exponential growth
occurs is modified slightly, but the rate of growth
(the slope) is approximately unmodified.
Whether or not there are oscillations of sign
depends on the rate at which lepton number is
being created during the exponential growth (small
changes in the value of $T_c$ would not matter).
For this reason, the approximation of integrating around
the resonance should be an acceptable approximation.

{\bf Figures 2a,b} are the same as figure 1a,b except
for the different choice of parameters,
$\delta m^2 = -100\ eV^2$, $\sin^2 2\theta_0 = 10^{-8}$.

As a final check of that static approximation
we numerically solve the QKEs for the 
entire momentum range ($0.1 < p/T < 12.0$) for one example.
This is quite CPU time consuming and it is most
economically done for small $\delta m^2$.
We take $\delta m^2 = -0.01\ eV^2$ and $\sin^2 2\theta_0 =
10^{-7}$.  For these parameters,
{\bf figure 3} compares the numerical solution of the QKE
with the static approximation, both integrated over the 
entire momentum range. Note that only the high temperature
evolution is shown. As the figure shows we obtain excellent
agreement as we expect
\footnote{In the appendix we give further details about
the approximation of integrating around the resonance.
We also compare the two different numerical procedures
employed.}.

Interestingly, Ref.\cite{dolgov} appears to have
(almost) rederived the static approximation 
(they neglect the chemical potentials in the
repopulation, which is important for the issue of sign).
They seem to get a quite different numerical solution
when they numerically solve the static approximation,
and are not able to obtain any solution at all
when they numerically solve the QKE's.
In our opinion, the remarkable agreement 
between our solution of the static approximation 
and the QKE's (as figures 1-3 illustrate) 
is a convincing check that we 
have done the numerical work competently. 
Of course this check has been done previously, and
there are examples already given in the literature\cite{fv2},
but since this work was ignored in Ref.\cite{dolgov},
we have taken the trouble to emphasise it again here.

\section{Statistical fluctuations}\label{sec:sf}

In this section we want to provide a simple 
estimation of the statistical fluctuations 
in the effective total lepton number. These arise 
simply because of the fluctuations 
in the number of particles within the region, 
around each oscillating neutrino wavepacket,
that determines the properties of the medium through 
the effective potential.

Let us first proceed without specifying the 
size of this region but we write it
in units of the interaction 
length ${\ell}_{int}\equiv \Gamma ^{-1}$ 
of the oscillating neutrino with $x\simeq 2.2$ 
(the peak of the distribution).
The number of photons within a 
region of size ${\ell}$ at the temperature $T$ is given by: 
\begin{equation}
N_{\gamma}\sim n_{\gamma}\,\ell ^3 \sim
10^{65}\,\left({{\rm MeV} \over T}\right)^{12} \, 
                  \left({\ell\over {\ell}_{int}}\right)^3.
\end{equation}
Since the number of neutrinos is of the same order
as the number of photons, it follows that the
statistical fluctuation of lepton number is of order:
\begin{equation}\label{eq:DeL}
\Delta L = {\sqrt{N_{\nu}}\over N_{\nu}} \sim
10^{-32}\,\left({T \over {\rm MeV}}\right)^6 \, 
      \left({\ell}\over {\ell}_{int}\right)^{-\frac{3}{2}}.
\end{equation}
Note that since $n_B \sim 10^{-10}n_{\gamma}$, it follows that
$\Delta \eta = \sqrt{n_B}/n_{\gamma} \sim 10^{-5}\Delta L$ and thus
$\Delta L^{(\alpha)} \sim \Delta L$.

We are interested to evaluate this 
fluctuation at the critical temperature $T_{c}$,
when the lepton number starts to be generated. 
The critical temperature can be put in 
the following approximated form\cite{fv1,pas}
\begin{equation}
T_{c}\approx 
\left({|\delta m^2|\over {\rm eV}^{2}}\right)^\frac{1}{6}
18\ {\rm MeV}.
\label{ttt}
\end{equation}
We can thus more usefully express the temperature in units 
of $T_{c}$ in Eq. (\ref{eq:DeL}), obtaining: 
\begin{equation}
\Delta L^{(\alpha)}\simeq 10^{-24}\,\left({T \over T_{c}}\right)^6 \,
			 {|\delta m^2|\over {\rm eV}^{2}}
                  \left({\ell}\over {\ell}_{int}\right)^{-\frac{3}{2}}.
\end{equation}

Now what should we take for the size $\ell$?
We will present a heuristic argument that 
$\ell \stackrel{>}{\sim} 10^{-2}\ell_{int}$. 
First of all, we find numerically that
neutrino asymmetries are typically not significantly
created on time scales less than of order $10^{-2}\ell_{int}$
i.e. $\ell_{int}{dL \over dt} \stackrel{<}{\sim} {\cal O}(10^2 L)$
for the parameter space of interest (this is true even in the 
exponential growth phase). This means that we only need
to consider {\it time} scales greater than $10^{-2}\ell_{int}$.
Furthermore since neutrinos are free streaming on such small
time scales it follows that the {\it distance} scale
for the creation of $L$ is greater than 
$10^{-2}\ell_{int}$. Thus the characteristic volume 
relevant for statistical fluctuations is 
larger than or of order $(10^{-2}\ell_{int})^3$.
Hence, for $|\delta m^2|\leq 1000 \ {\rm eV^2}$
we conclude that:
\begin{equation}
\Delta L^{(\alpha)} \stackrel{<}{\sim} {\cal O}\left(10^{-18}
\right).
\label{pop}
\end{equation}
In the next sections we will have to find out whether 
such fluctuations are able to induce 
a random sign of the final lepton number. 
Note that in addition to statistical fluctuations,
it is also possible to envisage inhomogeneities
which may arise from a model of lepto-baryogenesis for example.
Such inhomogeneities may potentially be much larger
than the statistical fluctuations but are of course model
dependent. Note that the possibility of such inhomogeneities 
has recently been considered in Ref.\cite{pas2} where
it is shown how the diffusion process must be taken into 
account to properly describe the dynamical evolution of
the neutrino asymmetry.

\section{Region with no oscillations}

In Ref.\cite{fv1} it was shown that in the framework 
of the static approximation the sign of the final 
value of total lepton number is the same as the sign of baryon 
asymmetry term,  $\tilde{L}$, and this is 
due to the presence of a 
correcting term into the equation
(see also Ref.\cite{pas2} for a detailed discussion). 
Furthermore, if the initial value of the effective
total lepton number 
is the same as that of $\tilde{L}$, then the effective total 
lepton number never changes sign during its evolution 
[otherwise it changes sign only once].
We will simply refer to this situation as 
{\em no oscillations}. 

In Ref.\cite{fv1,bvw} it was also shown that for small 
mixing angles and $|\delta m^2|\stackrel{>}{\sim} 
10^{-5} {\rm eV}^2$, 
one expects the static approximation to be valid.
In this section we present the results of a systematic 
research in the space of mixing 
parameters that confirm this expectation. This research 
has been done solving numerically
the QKE. The repopulation term has been taken 
into account in two different ways.

In the first one we simply describe the active neutrino 
distribution assuming a thermal
distribution ($f_{\nu_{\alpha}}=f_{eq}$).
 This assumption is equivalent to saying that 
the collisions are able to
instantaneously refill the states depleted 
by the oscillations. Such an assumption is
justified considering that for temperatures 
$T\gg 1 {\rm MeV}$ (which is typically the case
since the region where the sign is determined 
is during the period of
exponential growth which occurs
for $T_c \stackrel{>}{\sim} 3 \ MeV$
for $\delta m^2 \stackrel{>}{\sim} 10^{-5}\
eV^2$) and values of 
$s^{2}\ll 10^{-4}$,  one has respectively 
$\Gamma\gg H$ and $\Gamma\gg 
\Gamma_{\nu_{\alpha\rightarrow\nu_{s}}}$. 
In this case in the equations
(\ref{eq:b1}) it is approximately valid 
to replace $P_{0}$ and  $P_{z}$
with the neutrino distributions 
$z_\alpha \equiv f_{eq}/f_0$
and $z_s \equiv f_{\nu_{s}}/f_0$, via:
\begin{equation}
P_0 = z_\alpha + z_s,\ P_z = z_\alpha - z_s.
\end{equation}
The differential equation for $P_{z}$ gives a 
differential equation for
$z_{s}$
\begin{equation} 
\frac{dz_{s}(x)}{dt}=-\frac{1}{2\,}\beta(x)\,P_{y}(x).
\end{equation}
The differential equation for $P_{0}$ becomes redundant
because simply:
\begin{equation}\label{eq:dp0dt}
\frac{dP_{0}}{dt}=\frac{dz_{s}}{dt}+\frac{dz_{\alpha}}{dt}, 
\end{equation}
with $z_{\alpha}\equiv f_{eq}/f_{0}$, 
fully determined by the thermal equilibrium assumption 
\footnote{
A naive interpretation of the equation 
(\ref{eq:b1}) for $P_{0}$, would suggest that 
in the limit of thermal equilibrium $dP_{0}/dt=0$. 
This result is clearly incorrect because on the 
contrary it corresponds to the case that 
interactions are off and 
repopulation is not active at all. 
The assumption of thermal equilibrium 
physically corresponds to the case that 
the rate $\Gamma(x)$ of the collisions, 
that are refilling the states of momentum $x$, 
is so strong that the difference 
$f_{\nu_{\alpha}}-f_{\nu_{eq}}$ 
is very small. Thus it corresponds to the 
limit $\Gamma\rightarrow\infty$ 
with $dP_{0}/dt$ finite given by 
the Eq. (\ref{eq:dp0dt}), from which it follows that 
$f_{\nu_{\alpha}}\rightarrow f_{\nu_{eq}}$.}.

In the second way we calculated the repopulation
term using the repopulation
equation in Eq.(\ref{eq:b1}). 
The comparison shows nice agreement, 
as illustrated in  {\bf fig. 4a,b}.
Figure 4a is for the parameter choice
$\delta m^2 = -1\ eV^2, \ \sin^2 2\theta_0 = 10^{-7}$,
while figure 4b is for $\delta m^2 = -10\ eV^2, 
\ \sin^2 2\theta_0 = 10^{-7}$.
The solid line is the numerical integration
of the QKEs, Eq.(\ref{yd})
while the dashed line assumes the thermal
active neutrino distribution (discussed above).
Note that the two computations were done quite independently
using different codes.  
Also, two different initial values of $L^{(\tau)}$ were
chosen along with a slightly different initial temperature.
We have also done many other examples
which we do not show for environmental 
reasons (i.e. to save trees).
This confirms the expectations about the 
validity of a thermal equilibrium assumption.
Note that we only show the {\it high} temperature evolution
since this is the region where the sign is determined.
(Recall that the final magnitude of the neutrino
asymmetry is in the range $0.23 \stackrel{<}{\sim} L_{\nu_\alpha}
\ \simeq L^{(\alpha)}/2 \ 
\stackrel{<}{\sim} 0.37$ as illustrated in
figures 1a,2a and is discussed in detail in Ref.\cite{fv2}).

For fixed values of $|\delta m^2|$,
we studied the evolution of lepton number 
for increasing values of the mixing angle.
In the {\bf figure 5} we show an example 
$\delta m^2 = -10\, {\rm eV^2}$. It appears evident that for 
the shown values of mixing angles, the 
sign of lepton number does not oscillate.
As we will discuss later on, for small mixing
angles this result is expected. 
Of course changing the initial conditions does not change
this result as we illustrate in {\bf figure 6}
\footnote{
Of course qualitatively different behaviour does occur
if the initial value of $L_{\nu_{\tau}}$ is large
enough, roughly, $L_{\nu_{\tau}}^{initial} \stackrel{>}{\sim}
10^{-5}$\cite{ekm,prl}. For such a large initial value, the oscillations
cannot effectively destroy $L^{(\tau)}$ and $L^{(\tau)}$
remains of order $L_{\nu_{\tau}}^{initial}$ until lower
temperatures where it eventually further increases.}.
This figure considers $\nu_\tau \leftrightarrow \nu_s$
oscillations with $\delta m^2 = -10\ eV^2$ and $\sin^2 2\theta_0
= 10^{-7}$ with four different initial values of
$L_{\nu_\tau}$ employed.

The result of our careful study of the sign of
the neutrino asymmetry is given in {\bf fig. 7}.
As the figure shows, the parameter space
breaks up into two regions, a no oscillation region and
an oscillation region.  
Looking at the no oscillation region, one can easily
identify two different parts,
one part at low $\sin^2 2\theta_0$:
\begin{equation}
\sin^{2}\stackrel{<}{\sim}10^{-6}, 
\end{equation}
and one part at large $\sin^2 2\theta_0$,
\begin{equation}
\sin^{2} 2\theta_0\,\delta m^{2}\stackrel{>}{\sim} 3\times 10^{-4} \,eV^{2}.
\label{los}
\end{equation}
The existence of the no oscillation region at low $\sin^2 2\theta_0$ 
is consistent with the physical insight 
provided by the static approximation approach. 
Recall that in this approximation, oscillations of sign
do not occur \cite{fv1,pas}.
Furthermore this approximation is valid
for $\sin^2 2\theta_0$ sufficiently small.
This is because the rate of change of lepton number 
during the exponential growth epoch
increases proportionally with the value of $\sin^{2} 2\theta_0$. 

In the region of large $3\times 10^{-4} eV^2/\delta m^2
\stackrel{>}{\sim} \sin^2 2\theta_0 \stackrel{>}{\sim}
10^{-6}$ oscillations evidently
arise at the critical temperature. This is not 
inconsistent with the static approximation, 
since it is not valid in this region.
At the moment we do not have a clear analytic
description of the onset of rapid oscillations,
since the QKE's themselves do not provide much
physical insight. We can give however a qualitative partial
explanation noticing that for $a \ll b$ [recall that $a,b$ are defined in
Eq.(\ref{sal})],
\begin{equation}
\left. \beta/D\right|_{res} \simeq 50 \sin2\theta_0. 
\end{equation}
Thus, the line $\sin^2 2\theta_0 \approx$ constant  ($10^{-6}$)
also corresponds to $\beta/D \approx$ constant ($10^{-1}$) at the 
resonance. At the resonance the quantity $\beta/D$ gives approximately
the value of the ratio between the mean interaction length and 
the matter oscillation length. Until this quantity is much less than unity
the coherent behaviour of neutrino oscillations at the resonance is averaged out
by the effect of collisions and the static approximation provides
a good description for the evolution of the neutrino asymmetry. 
When this quantity starts to approach unity this starts to be not true any more.
Thus, during the initial exponential 
growth, it is possible that some small effect
due to the oscillations between collisions, which
is an effect not included in the static approximation,
may be responsible for the rapid oscillations of the neutrino asymmetry.

Of course one may wonder why the oscillation region
stops when the mixing angle becomes large
enough, Eq.(\ref{los}). 
Actually there is quite a simple
explanation for this result.
For large mixing angles such as these, it is known that
the exponential generation of lepton number is
delayed because of the sterile neutrino production\cite{fv1,pas}.
At the same time, the initial exponential growth is considerably
weaker. A consequence of the initial slow exponential
growth of lepton number the static approximation is valid and
oscillations in the sign of lepton number cannot occur while
the neutrino asymmetry is growing relatively slowly. 
Thus this seems to explain why the
no oscillation region can extend also to large mixing angles.
Infact, in {\bf fig. 8} we show how for 
$|\delta m^{2}|=10^{2.7}\,{\rm eV}^{2}$ 
the critical temperature is delayed  and the 
rate of growth decreased just when the oscillations 
would be expected to appear at $s^{2}\simeq 10^{-6}$.
This is surely the reason why for
$|\delta m^{2}|\stackrel{>}{\sim} 10^{2.7}\,{\rm eV}^{2}$,
the oscillations do not occur for any value of mixing angle.
On the other hand for a fixed value of 
$|\delta m^{2}|\stackrel{<}{\sim} 10^{2.7}\,{\rm eV}^{2}$
the oscillations can occur between $s^{2}\simeq 10^{-6}$ and 
$s^{2}\simeq 3\times 10^{-4}\,({\rm eV^{2}}/|\delta m^{2}|)$.
 
We have now to discuss whether the presence of
statistical fluctuations with the amplitude estimated 
in section \ref{sec:sf}, can give any effect in the
``no oscillations" region.

In the static approximation framework, one can 
show the existence of an approximate fixed point, that for $T>T_{c}$ 
is stable and drives the effective total lepton number 
- $L^{(\alpha)}$ to values 
of order $10^{-15}-10^{-16}$ prior to the onset of the 
instability (see \cite{pas2}). This is clearly 
confirmed looking at the solutions in {\bf figures 1-6} 
and {\bf 8}, obtained solving directly the QKEs.

Therefore, from our estimation of the 
maximum amplitude of the statistical fluctuations,
Eq.(\ref{pop}), we can safely exclude any influence of them 
in determining the final sign of lepton number. 
We can thus conclude that {\em for values of mixing 
parameters in the region where the lepton number 
does not oscillate, given a certain fixed value 
of $\tilde{L}$, the final sign of lepton number 
is the same as that of $\tilde{L}$ and cannot randomly 
change in different spatial points of the early universe}.

In {\bf fig. 7} we have also showed the contour (dot-dashed line 
plus the dotted line) of the allowed region for the 
solution $\nu_{\mu}-\nu_{s}$ to the atmospheric neutrino anomaly, 
in the 3 neutrino mixing mechanism proposed in \cite{fv1} 
and reanalyzed in \cite{f,pas}. 
Even assuming that an 
overproduction of sterile neutrinos would occur 
in the parameter region of rapid oscillations, 
as claimed in \cite{shifu}, it appears evident that 
{\em the mechanism can still allow acceptable
values of the tau neutrino mass and is not
inconsistent with standard BBN}.

We want now to investigate more in detail
the nature of the rapid oscillations.

\section{Region of rapid oscillations}

The study of the rapid oscillation region is much 
harder from a numerical point of view. It means that to 
describe it properly the 
numerical account of momentum dependence
must be much more accurate (on this point
more details are given in the numerical Appendix). 
In particular a much higher resolution in momentum
space is necessary to describe the distributions.
This makes difficult a systematic exploration
of this region for different values of mixing parameters.

We mainly performed the runs for $\delta m^{2}= -10\, {\rm eV^{2}}$,
a particularly interesting value for the three neutrino mixing
mechanism mentioned at the end of previous section. In fact it would 
correspond to minimum values of the tau neutrino mass  
included in the allowed region and compatible with
structure formation arguments.

The transition from the region with no oscillations to that one with
rapid oscillations is very sharp, meaning that the width of the
boundary between the two regions is very narrow. 
In {\bf figure 9} we give an example of this transition 
for a fixed value of $\delta m^{2}=-10\, {\rm eV}^{2}$. 
Changing the value of $\log(\sin^{2}\theta_{0})$ for a 
very small quantity, $\sim 0.1$, is sufficient to pass from one
region to another. 

The oscillations take place in a limited interval of 
temperatures. This interval increases going from points at 
the border to points in the middle of the 
rapid oscillations region in the mixing parameter space. 
This feature
again confirms the expectation that the oscillations
are the effect of the breaking of the adiabaticity condition
during some period of the lepton number growth and that
moving toward the center of the rapid oscillations region, 
this period becomes longer.

A first observation about the rapid oscillations is 
that they develop with a temperature period 
of about $(10^{-3}-10^{-2})\,{\rm MeV}$, depending 
on the value of $|\delta m^{2}|$.
This period is roughly described by the
interval of temperature corresponding to 
the interaction mean time of neutrinos $\Gamma^{-1}$ 
for $x\simeq 2.2$:
\begin{equation}
\Delta_{int}T=\frac{H\,T}{\Gamma}\simeq 2\,{\rm MeV}\,
	\left(\frac{\rm MeV}{T}\right)^{2}
\simeq 5\times 10^{-3}\left(\frac{|\delta m^{2}|}{\rm eV^{2}} \right)
^{-\frac{1}{3}}{\rm MeV}.
\end{equation}
This is in agreement with the expectation for which
the coherent nature of the oscillations is damped by the 
collisions unless the effective potential is rapidly changing.

In the numerical appendix we will show how increasing the 
resolution in the momentum space for the description of
the distributions, has the effect of diminishing the amplitude 
of the oscillations. For the example in figure 9
with $\sin^{2}\,\theta_{0}=\,10^{-6.16}$, this effect 
is such that if the resolution is not high enough, then
the amplitude of the oscillations is so large that
the total lepton number {\em changes sign}, while when
a good resolution is employed and the numerical 
solution becomes stable with changing the resolution,
the sign changes disappear. If the oscillations are artificially 
amplified by a numerical error, this can induce the suspect
that  the sign changes are not a real feature
of the solutions. However for a choice of mixing
parameters more inside the rapid oscillations 
region, like already for $\sin^{2}\,\theta_{0}=\,10^{-6.1}$
in figure 9, it seems that the sign changes do not disappear
increasing the accuracy. As we will say in more detail 
in the appendix, on this point we cannot be 
conclusive because we cannot reach the required 
resolution to have perfect numerical stability.
It seems that more advanced computational means are
necessary to get a definitive conclusion, while in this work we 
prefer to say that there is an {\it indication} 
that the lepton number oscillations change the sign of the
solution during the growth regime.
We think however that on this point it would be desirable to 
have some rigorous analytical demonstration,
that in our opinion is still missing, even though
some attempts have been recently done \cite{Sorri99}.

Another important problem is whether the final sign of lepton 
number is sensitive or not to the small statistical fluctuations
necessarily present in the early Universe, as we saw in the
third section. In our numerical calculations we observe 
that changes in the initial
lepton number are able to change the final sign with a choice of 
mixing parameters well inside the rapid oscillations region
where a large number of oscillations take place. While
moving toward the border this effect tends to disappear.
This behaviour seems to confirm an effect of `dephasing' of the solutions
as described in \cite{Sorri99}. This means that even though solutions
with different initial conditions converge at the same values prior
the onset of the instability at the critical temperature,
(as shown for example in the figure 6), small relic differences
will afterwards develop, resulting in growing phase differences 
during the oscillatory regime and in a possible final sign inversion.
However also on this point we cannot be conclusive.
It is infact necessary first to get a
perfect numerical stability in the description of the
neutrino asymmetry oscillations and then to prove that
statistical fluctuations as tiny as those present in the
early Universe are able to yield a dephasing effect.
For this second step it is then necessary that the numerical error
is lower than the statistical fluctuations.
Thus we cannot be conclusive because we cannot reach such 
a computational accuracy. This means also that our analysis cannot 
neither exclude or demonstrate the possibility of a 
chaotical generation of lepton domains, 
but in any case we can say that this possibility is limited to a region of
mixing parameters contained in the rapid oscillations region, 
in a way that the neutrino asymmetry undergoes a number of oscillations 
high enough that two initially close values of the neutrino asymmetry, 
result, in the end, in a different final sign. The possibility
of drawing the exact contour of such a region, inside the rapid 
oscillations one, is again something
that requires a numerical analysis beyond our present
computational means.
 
We can thus conclude this section saying that the expectation of 
the appearance of coherent effects on the evolution of lepton number
when an adiabatic condition breaks is confirmed by the numerical
analysis. This suggests that the oscillations are not just an effect of 
the numerical error but an intrinsic feature of the equations.
The amplitude of these oscillations is much likely large enough
to change the sign of the neutrino asymmetry, but on this
point we cannot be conclusive, because our maximum 
accuracy in the description of the momentum dependence is not 
sufficient to get a good numerical stability and we prefer
to speak of indication of sign changes.
The demonstration that the changes of sign can induce a chaotical
generation of lepton domains is beyond our numerical analysis,
but in anycase this can occur only in a region confined within
the region of rapid oscillations shown in figure 7.
Therefore any application that makes use of a chaotical
generation of lepton domains as a general consequence of the QKE,
is not justified. 

\section{Conclusions}

Large neutrino asymmetry is generated in
the early Universe by ordinary -
sterile neutrino oscillations for a large range 
of parameters. We have 
made a numerical study of the final sign
of this asymmetry and our results are summarized in figure 7.
In the space of mixing parameters we identified a no oscillation
and an oscillation region. 
In the no oscillation region we could show perfect
agreement with the predictions of the static approximation
for temperatures $T\stackrel{>}{\sim} T_c$ at which the MSW
effect is inhibited by the presence of the collisions.
Moreover perfect numerical stability of the solutions was
obtained. These two facts do not leave any room for
doubt about the generation of a neutrino asymmetry and
its final value.

We have also estimated an upper limit [see Eq.(\ref{pop})] 
for the size of statistical fluctuations of the
lepton number which is very tiny.
In view of this, for a choice of mixing parameters in the 
no oscillation region, the final sign
of the neutrino asymmetry is the same in all
points of space unless there happens to be larger fluctuations 
(of non-statistical origin) present.
In this way the possibility of a
chaotical generation of lepton domains 
is excluded in the no oscillation region. 
One consequence of this is that
the three neutrino mixing mechanism (proposed in \cite{fv1}) allowing 
the $\nu_{\mu}\leftrightarrow\nu_{s}$
solution to the atmospheric neutrino anomaly to be
consistent with a stringent BBN bound of $N_{BBN}^{eff} < 3.5$ 
is still viable.

We have also been able to plausibly
justify the existence of the rapid oscillations region
in terms of a  breaking of the static approximation
during the lepton number growth.

The study of the rapid oscillations in the oscillation region
encounters many numerical
difficulties. The oscillations seem to have an amplitude sufficiently 
large to change the sign of the total lepton number during the
growth regime. We however think that further analysis are required
to be conclusive on this point. 
We also think that at the present the numerical analysis 
cannot neither exclude or demonstrate the possibility of a 
chaotical generation of lepton domains. 
This is because the error introduced by the
numerical analysis is much larger than
the tiny statistical fluctuations present in the early universe.
In anycase this possibility is confined to a special choice 
of the mixing parameters, corresponding to some region
contained within the region of rapid oscillations.

Finally we should also mention that while we have focussed on
$\nu_\alpha \leftrightarrow \nu_s$ oscillations in isolation,
our results will be applicable to realistic models with sterile
neutrinos. In the more general case of three ordinary neutrinos
and one or more sterile neutrinos, the neutrino asymmetry generation
occurs first for the oscillations 
with the largest $|\delta m^2|$ (with $\delta m^2 < 0$).
The sign of this asymmetry (which we assume to be
$L_{\nu_\tau}$ for definiteness) will be the same as $\tilde{L}$ except
possibly in the oscillation region where it may depend on 
$\delta m^2,\ \sin^2 2\theta_0$. The other oscillations with
smaller $|\delta m^2|$ may also generate significant neutrino
asymmetry, although this generation occurs latter on (see
e.g. Ref.\cite{fv2,nfb} for some examples). 
The sign of these asymmetries will ultimately 
depend only on the sign of $L_{\nu_\tau}$ since this dominates 
$\tilde{L}$ for these oscillations.

\section*{Numerical Appendix}

In this Appendix we want to describe in more detail our numerical
procedure for solving
the QKE's. As we said already in the text 
we employed two different codes, one using the 
thermal equilibrium assumption ({\em code A}), the other using the expression
(\ref{eq:b1}) for the repopulation account ({\em code B}). The 
code B has been already described in \cite{f} and we refer the reader 
to this reference for details. Here we describe common features 
of the two and specifications of the code A. 
 
We first discuss the time step and then the momentum
integration is described.
 
The time step of integration is adjusted in a way that it is halved
until the required accuracy (the relative error on all the set 
of variables of the system) $\epsilon$ is reached.
By this we mean that for each variable 
we introduce a time step
$\Delta t$. At every time step during
the evolution, say at $t=t_x$ we compute
$Z(t_x + \Delta t)$ and compare this with
the results of integration with a half step size,
symbolically, $Z(t_x + \Delta t/2 + \Delta t/2)$.
Explicitly, the step size is halved until 
every integration variable (which
we have denoted collectively as $Z$) satisfies,
\begin{equation}
|Z(t_x + \Delta t) - Z(t_x + \Delta t/2 + \Delta t/2)|
< \epsilon 
|Z(t_x + \Delta t) + Z(t_x + \Delta t/2 + \Delta t/2)|/2.
\end{equation}
The above procedure is quite standard, and
is called Merson' s procedure' 
or {\em step doubling} \cite{lan,numrec}
and together with a fourth order Runge-Kutta solver is
considered the most straightforward and safe technique
\footnote{
An evolution of this procedure is given by
the ``Fehlberg" or {\em embedded Runge-Kutta method}.
This method should be twice as fast as Merson's one
and although it should be safe enough, we have
adoped the more conservative Merson's procedure for code A.
Another possibility is offered by the 
Bulirsch-Stoer method that is by far the most efficient
but it has many cautions for its usage. 
Same considerations hold for predictor-corrector methods 
that are considered somehow worse than all of the previous ones.
As we had to study unexplored features of 
still `young' equations, 
we preferred to adopt the {\em step doubling} method
(see \cite{numrec} for further details).}.

We now discuss the discretization of momentum and
the integration around the resonances procedure.
The range of values of momenta ($p/T$)
is $[x_1,x_N]$ which is discretized 
on a logarithmically spaced mesh:
\begin{equation}
x_ j= x_{1}\times 10^{(j-1)\,\Delta} , \ j = 1,2,...,N
\end{equation}
where $N$ is the number of bins and $\Delta\equiv\log(x_{N}/x_{1})/(N-1)$. 
The initial temperature is chosen in a way that the 
initial resonant momentum is $x_{in}$.

The resonance momentum for neutrinos, $x_{res} \equiv p_{res}/T$ 
can be obtained by solving $\lambda (x) = 0$ [with
$\lambda (x) $ defined in Eq.(\ref{sf})] and is given by 
\begin{equation}
x_{res} = {X_2 \over 2X_1} + \sqrt{\left({X_2 \over 2X_1}\right)^2
+ {c \over X_1}},
\end{equation}
where $c \equiv \cos 2\theta_0$ and
$X_1 \equiv b/x^2, X_2 \equiv a/x$ 
[recall $a,b$ are defined in Eq.(\ref{sal})].
Note that $X_{1,2}$ are independent of $x$.
The resonance momentum for antineutrinos can
be obtained by replacing $X_2 \to -X_2$ in the above equation.
The width of the resonance is given by the following expression
\footnote{This is derived in Ref.\cite{f}. The form given
in Eq.(\ref{xyy}) is simpler, but mathematically
equivalent to $\Delta/p_{res}$ where 
$\Delta$ is given in Eq.(28) of Ref.\cite{f}.}:
\begin{equation}
\delta x=2\,\left[\frac{\sqrt{\sin^2 2\theta_0 + 
d_{0}^{2}\,b^{2}}}{c+b}\right]_{x_{res}}\,x_{res}
		\equiv \delta_{res}\,x_{res}, 
\label{xyy}
\end{equation}
where we introduced the {\em logarithmic width of the resonance} 
$\delta_{res}=\delta\ln x$ and
$d_{0}\equiv 2p_{res}D/b\delta m^2$. Numerically, 
$d_0 \simeq 2\,(0.8)\times 10^{-2}$, for $\alpha=\mu,\tau$ ($e$).
Note that $\delta_{\rm res}$ is roughly independent of
$T$ for $T \stackrel{>}{\sim} T_c$ (and is of order $d_0$)
while for $T \ll T_c$, $\delta_{\rm res}$ decreases to
$\sim \sin2\theta_0$.
It is also roughly independent of the momentum $x$
and this is the reason for which it is more convenient to
use a logarithmically spaced mesh rather than a linearly
spaced one.

At each step of integration we calculate both $x_{\rm res}$ and
$\delta_{\rm res}$ and we discretize them: $x_{\rm res}\rightarrow 
j_{\rm res}$
with
\begin{equation}
j_{\rm res} = 
Int\left[ {1 \over \Delta}log{x_{\rm res}\over x_1}\right]+1
\end{equation}
and we define $\Delta j\equiv 1+{\rm Int}(\delta_{\rm res}/\Delta)$.
Considering that $\delta_{\rm res}=\delta (\ln x)
\simeq 2.3\times\delta\log x$, this means that $\Delta j$ corresponds
roughly to twice the width of the resonance in a logarithmic scale and in
units of $\Delta$. At each time step we integrate only
on those bins among $[1,N]$ in the interval 
$[j_{\rm min}, j_{\rm max}]$ where:
\begin{equation}
j_{\rm min}=j_{\rm res}-\rho\, \Delta j,\hspace{3mm}
j_{\rm max}=j_{\rm res}+\rho\, \Delta j
\end{equation}
In this way we are integrating symmetrically around the resonance 
(we are taking equal numbers of bins with lower and higher momenta), but 
logarithmically, while in the code B this is done linearly \cite{f}. 
Clearly $j_{\rm min}$ and $j_{\rm max}$ are constrained within
$[1,N]$.
This procedure is done at each step so that
the interval of integration follows dynamically the resonance.
Initially the neutrino and anti-neutrino resonances approximately
coincide due to the approximate fixed point $L^{(\alpha)} \approx 0$.
However during the exponential growth the resonance splits in two,
one at low momenta for the antineutrinos and one at high momenta for
the neutrinos if $L^{(\alpha)} > 0$ or vice versa if $L^{(\alpha)} < 0$. 
The split occurs gradually: first the initial interval enlarges and at 
certain point it splits, so that eventually the two intervals do not overlap.

Let us introduce the number of bins which we
integrate over, which for neutrinos
we denote by $M_\nu$ and is
given by $M_{\nu} \equiv \min(N,J_{\rm max}) - \max(1,J_{\rm min})$
(and similarly for antineutrinos we introduce the quantity $M_{\bar{\nu}}$).
So at each step of integration we solve a system of
$3(M_{\nu} + M_{\bar \nu})+1$  equations in the code A
and $4(M_{\nu}+ M_{\bar \nu})+1$ in the code B.

In this way one has $6$ numerical parameters: 
$\epsilon, x_{1}, x_{N}, x_{in}, \rho, N$.
 One has to check carefully that the numerical 
solution is `stable' changing these parameters in the direction
of greater accuracy.

For the parameter $\epsilon$ we checked 
that $\epsilon=10^{-4}$ is a safe value.
A safe choice for 
$(x_{in}, x_{1}, x_{N})$ is $(0.1, 10^{-3}, 20)$
[$x_{in}=0.1$ corresponds approximately to the 
choice $T_{in}\simeq 3\times T_{c}$ 
where $T_c$ is given by Eq.(\ref{ttt})].

The minimum value for $N$ to get an accurate description, 
depends on the values of the mixing parameters,
whether they are outside or inside the rapid oscillations region.
The number of bins $N$, necessary to have a stable solution, is that
one for which the number of bins within a width of resonance,
$\Delta j$,   
is large enough. A value $\Delta j\sim 10$, 
that corresponds to $N\sim 2^{11}$, is sufficient in the no
oscillatory region to have stable solutions as shown
in {\bf figure 10a}. 

A first observation concerning the value of $\rho$ to be chosen
is that $\rho \sim 100$  corresponds roughly to integrate on 
all the range of momenta $10^{-3}-20\;$
\footnote{If we start from $x_{in}=0.1$ and if we consider 
the case $\alpha=\mu, \tau$ so that $\delta_{res}\simeq d_{0}\simeq 0.02$ 
(for $T\stackrel{>}{\sim} T_c$), then the choice $\rho=100$ is
equivalent to integrating on the interval
$10^{-3} < x < 10$.
At the critical temperature $x_{res}\approx 2$, the integration
interval moves at values $0.02 < x < 20$. 
This means that the integration
around the resonance is not necessarily more approximated than
the case when a {\em static} interval of integration is chosen,
it is simply a way to select a reduced interval of momenta but
dynamically, taking into account the resonant behaviour of the
process.}.
For values of mixing parameters in the no oscillation region, 
$\rho=4$ is a good choice in order to study the problem of sign.  
In {\bf figures 10a, 10b} 
we present the result of a check that shows the 
numerical stability of our results for the indicated
values of the mixing parameters. In figure 10a the 
mixing parameters
correspond to a point close to the border with  
the rapid oscillations region.

More precisely the minimum value of $\rho$ to get the stability, 
in the no oscillations region, depends on the initial value of 
the total lepton number. If one choose $L^{(\alpha)}_{in}=0$, a relative
low value of $\rho=4$ is sufficient to get a full stability at all
temperatures in an impressive way. On the other hand if one starts from 
non vanishing values of the initial lepton number, the 
approximation of integrating
around the resonance gives an error 
at high temperatures, in the regime during which 
the lepton number is destroyed. This is 
also clearly shown in figure 10a and 10b for the indicated mixing parameters.
This fact has already been described and justified in the main text. 

In the oscillatory region, in order to get stability, the 
necessary number of bins inside a resonance width
seems to hugely increase upto about $\Delta j\simeq 500$,
obtained for $N\simeq 2^{17}$, as shown in {\bf figures 11a, 11b}.
It is remarkable to notice how an inaccurate description of
the momentum dependence amplifies the amplitude of the oscillations
and for $\sin^{2}\,2\theta_{0}=\,10^{-6.16}$ it makes the solution 
changing sign. However for an higher value of the mixing angle the 
sign changes are present also when the maximum accuracy is employed.
The stability of the solutions  in the oscillatory regime
is clearly not as good as outside this regime, even though there is still
some indication on the behaviour of the solutions.
Not only it would be desirable to make checks increasing the number of bins $N$,
but also increasing the width of the region of integration around the
resonance, that means the parameter $\rho$. Unfortunately this
is not possible not only because the required computational time becomes too 
long, but also because the dimensions of the arrays
employed to describe the distributions requires an amount
of RAM memory than our machines do not have. It would be also desirable
to make checks increasing the numerical precision, in our case passing
from a double precision (the round off error is about $10^{-16}$)
to a quadruple precision. This because at the onset of the
instability the rate of growth of neutrino asymmetry is the
difference of two opposite quantities that are much higher.
Unfortunately we do not have at the moment the possibility
to perform quadruple precision runs, also because again the 
time for the runs becomes too long. For all these reasons we prefer 
to speak of  a strong `indication' that the oscillations are able   
to change the sign of the neutrino asymmetry.

In the end of this appendix we want to mention that
also for an accurate evolution at low temperatures $T \ll T_c$,
large values of $N$ are required (depending on $\sin2\theta_0$).
This is because, the resonance width becomes quite narrow 
at low temperatures [see the discussion 
above, following Eq.(\ref{xyy})].

This appendix has been stimulated by the scepticism 
expressed in \cite{dolgov} 
about the numerical procedure employed in the previous existing works 
dealing with the numerical solutions of the exact QKE. 
It should be clear now that the results and the 
procedure were already rigorous
at that time, just there was not a compelling necessity 
to provide all the tedious details as we do now. 

\vskip 0.4cm
{\bf Acknowledgements}
\vskip 0.3cm
We thank Nicole Bell, Roland Crocker, Ray Volkas and Yvonne Wong
for valuable discussions.
P. Di Bari wishes to thank Maurizio Lusignoli for helpful discussions
and for his support and encouragement to the collaboration with
R. Foot and the Melbourne University.
He also wants to thank Piero Cipriani, Maria Di Bari, Maurizio Goretti,
Trevor Hales and
Rob Scholten for help with
his portable computer making his work in Melbourne much easier and faster.
R.F. acknowledges the kind hospitality of Paolo Lipari and Maurizio
Lusignoli at the University of Rome during a visit where
this work was initiated.


\section*{Figure captions}

{\bf Fig. 1} 
Evolution of the effective total lepton number, $L^{(\tau)}$,
for $\nu_\tau - \nu_s$ oscillations  
for the parameter choice, $\delta m^2 = -10\ eV^2$,
$\sin^2 2\theta_0 = 10^{-7}$. In Fig.1a,
the solid line is the numerical solution of the QKEs,
Eq.(\ref{yd}) integrating around the resonances
using $f = 12$ resonance widths and
in a linearly symmetric way
(the `code B' has been used: see the Appendix
and \cite{f} for its definition), while the dashed line
is the static approximation (described
in the text) also integrating around the resonance.
Fig.1b is a comparison of the static approximation 
integrating around the resonance (solid line)
with integrating over the entire momentum range $0.01 < p/T < 20$
(dashed line).
Note that $\tilde{L} = 5\times 10^{-10}$ is used
and the initial value of $L_{\nu_{\tau}}$
is zero in this example.

{\bf Fig.2a,b}
Same as figures 1a,b except that the
oscillation parameters are changed to
$\delta m^2 = -100\
eV^2$, $\sin^2 2\theta_0 = 10^{-8}$.

{\bf Fig. 3}
Numerical solution of the QKE's (solid line) integrated over the 
entire momentum range $0.1 < x < 12$ for the parameter choice
$\delta m^2 = -0.01\ eV^2, \ \sin^2 2\theta_0 = 10^{-7}$.
The initial value of $L_{\nu_\tau}$ is zero in this example.
Also shown is the static approximation integrated over the
same momentum range and same initial conditions (dashed line).

{\bf Fig. 4a,b}  
High temperature
evolution of the effective total lepton number,
$L^{(\tau)}$, for $\nu_\tau - \nu_s$ oscillations.
Figure 4a is
for the parameter choice, $\delta m^2 = -1\ eV^2$,
$\sin^2 2\theta_0 = 10^{-7}$ 
(with $L^{(\tau)}_{initial} = \tilde{L} = 5\times 10^{-10}$,
i.e. initial $L_{\nu_\tau} = 0$)
while figure 4b is for
$\delta m^2 = -10 \ eV^2$, $\sin^2 2\theta_0 = 10^{-7}$
(with $L^{(\tau)}_{initial} = 10^{-12}$).
The solid line is the numerical solution
of the QKEs, Eq.(\ref{yd}), while the dashed
line is the solution of the QKEs assuming a thermal 
distribution for the active neutrino (as discussed in
the text).

{\bf Fig. 5}
An example of the high temperature evolution
of the effective total lepton number, $L^{(\tau)}$
for $\delta m^2 = -10\ eV^2$ for $\sin^2 2\theta_0 =
10^{-9}$ (dotted line), $10^{-8}$ (dashed line)
$10^{-7}$ (solid line).

{\bf Fig. 6}
An example of the high temperature evolution
of the effective total lepton number, $L^{(\tau)}$
for $\delta m^2 = -10\ eV^2$ for $\sin^2 2\theta_0 =
10^{-7}$ with the initial values of $L_{\nu_\tau}$ given
by $L_{\nu_\tau} = 10^{-10}$ (solid line), $10^{-9}$ (long 
dashed line), $10^{-8}$ (short dashed line), $10^{-7}$ (dotted 
line).

{\bf Fig. 7}
Region of parameter space where the final sign 
does and does not oscillate for $\nu_\tau \leftrightarrow
\nu_s$ oscillations.
Also shown is the `allowed region' (which we define below)
for the $\nu_\mu \leftrightarrow \nu_s$ maximal oscillaton
solution to the atmospheric neutrino anomaly.
This is the region where the $L_{\nu_\tau}$ is
generated rapidly enough so that the sterile neutrino
is not equilibrated by either $\nu_\mu \leftrightarrow
\nu_s $ maximal oscillations (region above the dashed-dotted line)
or by $\nu_\tau \leftrightarrow \nu_s$ oscillations (region
to the left of the dotted line).

{\bf Fig. 8} The evolution of the total lepton number 
for a choice of mixing parameters just outside the top 
of the rapid oscillations region. The sterile neutrino production 
increases with the mixing angle.  It appears clear 
the double effect of the sterile neutrino production
in delaying the onset of the growth, lowering the critical 
temperature and in depressing the rate of growth of lepton 
number. The latter prevents the arise of rapid oscillations
when the mixing angle increases. 

{\bf Fig. 9}. Transition from the no oscillations region
to the rapid oscillation region increasing the value of the 
mixing angle for a fixed value of $\delta m^{2}= -10\ \rm{eV^{2}}$.

{\bf Fig. 10} Check of numerical stability for two different 
choices of mixing parameters inside the no oscillation 
region for $\delta m^{2}=-10\ {\rm eV^{2}}$ and 
$\sin^{2}\,2\theta_{0}=\,10^{-6.2}$ (a),
$\sin^{2}\,2\theta_{0}=\,10^{-7}$ (b). 
It appears clear the validity of the resonant 
approximation in this region when $L^{(\alpha)}_{\rm in}=0$.
On the other hand for 
$L^{(\alpha)}_{in}=\tilde{L}=5\times 10^{-11}$, 
the resonant approximation produces an error in 
the regime of high temperatures ($T\gg T_{c}$) 
when the initial lepton number is destroyed.

{\bf Fig. 11} Check of numerical stability in the
oscillatory regime for  $\delta m^{2}=-10\ {\rm eV^{2}}$
and $\sin^{2}\,2\theta_{0}=\,10^{-6.16}$ (a),
$\sin^{2}\,2\theta_{0}=\,10^{-6.2}$ (b). 

\newpage
\epsfig{file=xF1a.eps,width=15cm}
\newpage
\epsfig{file=xF1b.eps,width=15cm}
\newpage
\epsfig{file=xF2a.eps,width=15cm}
\newpage
\epsfig{file=xF2b.eps,width=15cm}
\newpage
\epsfig{file=xF3.eps,width=15cm}
\newpage
\epsfig{file=xF4a.eps,width=15cm}
\newpage
\epsfig{file=xF4b.eps,width=15cm}
\newpage
\epsfig{file=xF5.eps,width=15cm}
\newpage
\epsfig{file=xF6.eps,width=15cm}
\newpage
\epsfig{file=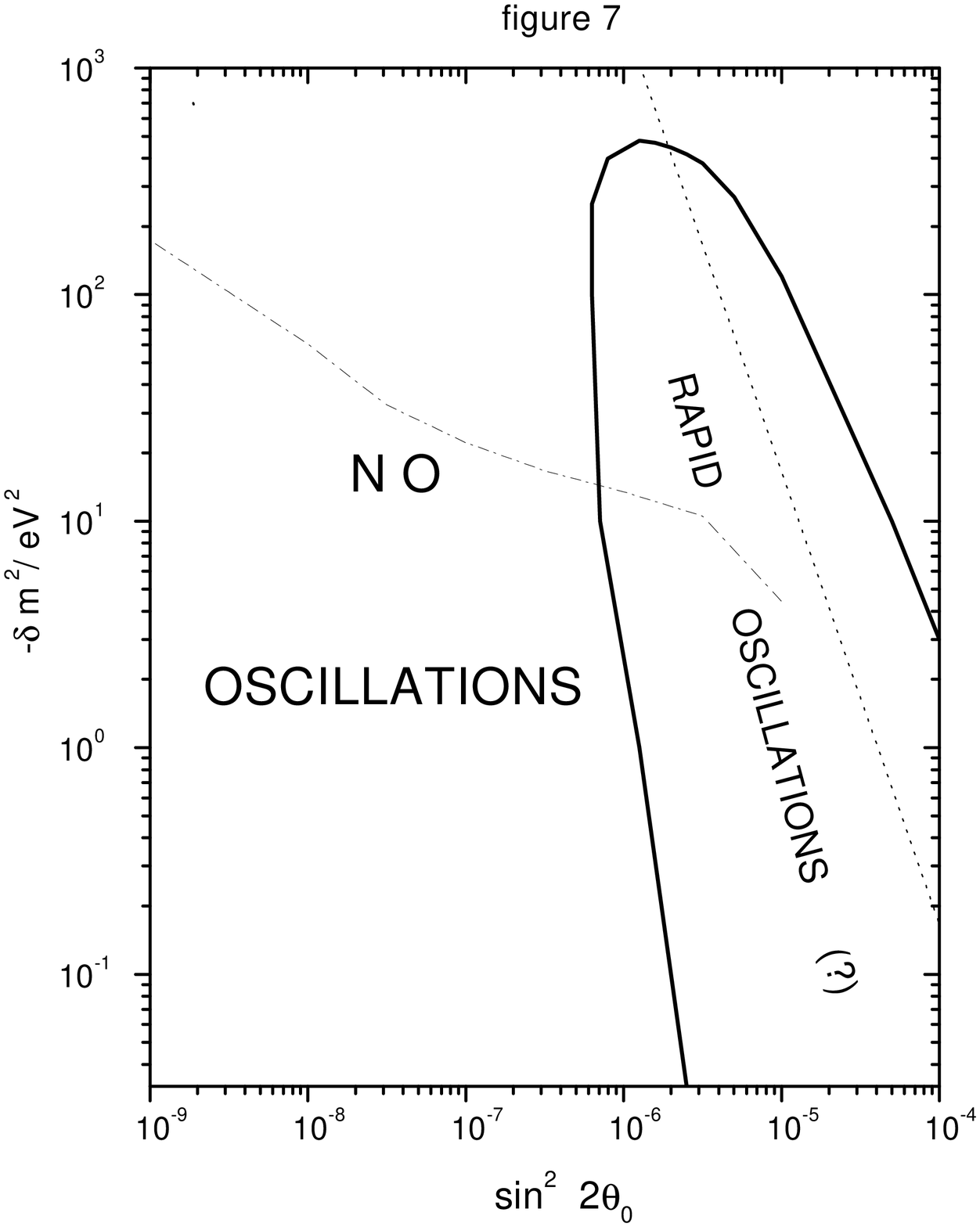,width=15cm}
\newpage
\epsfig{file=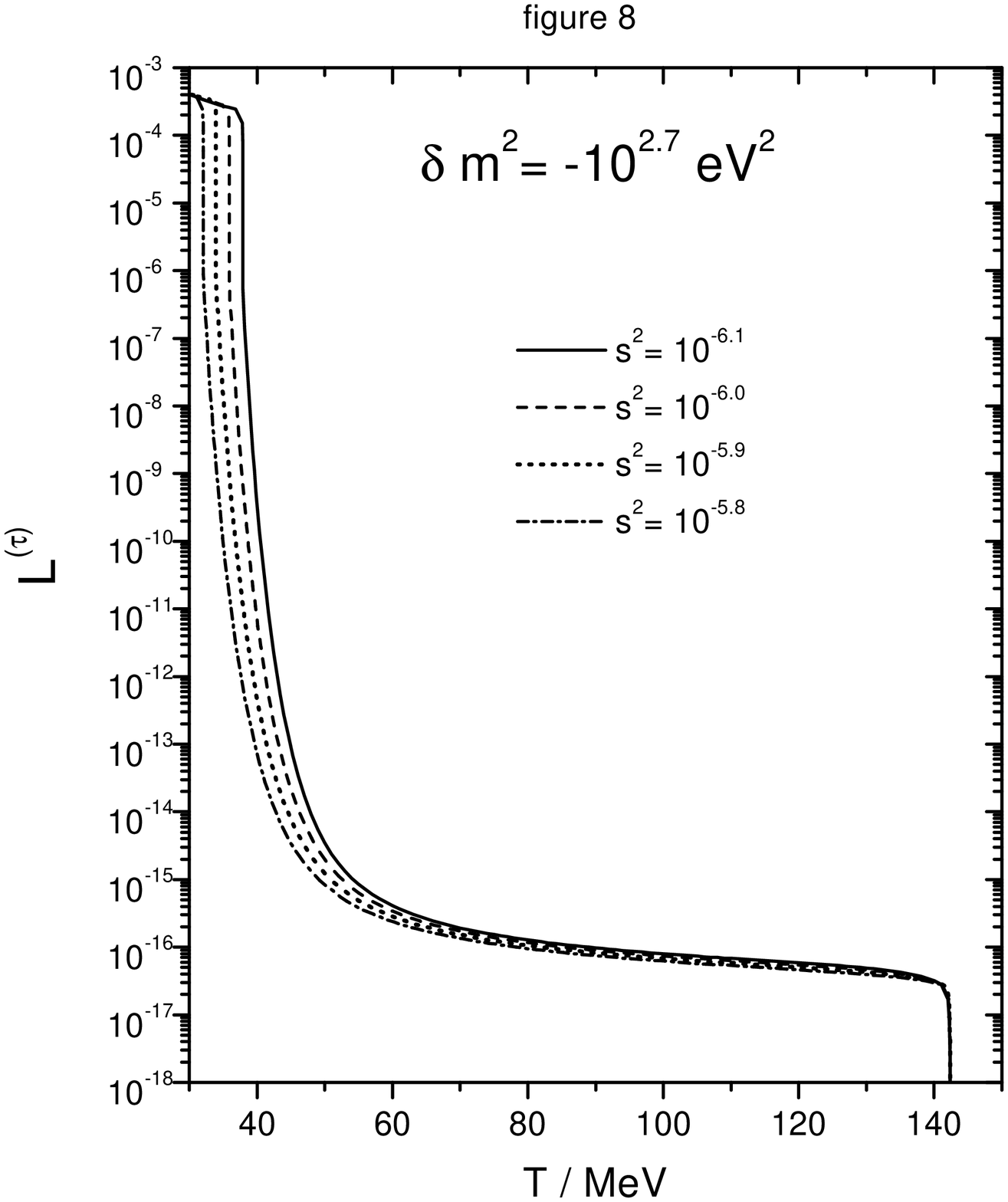,width=15cm}
\newpage
\epsfig{file=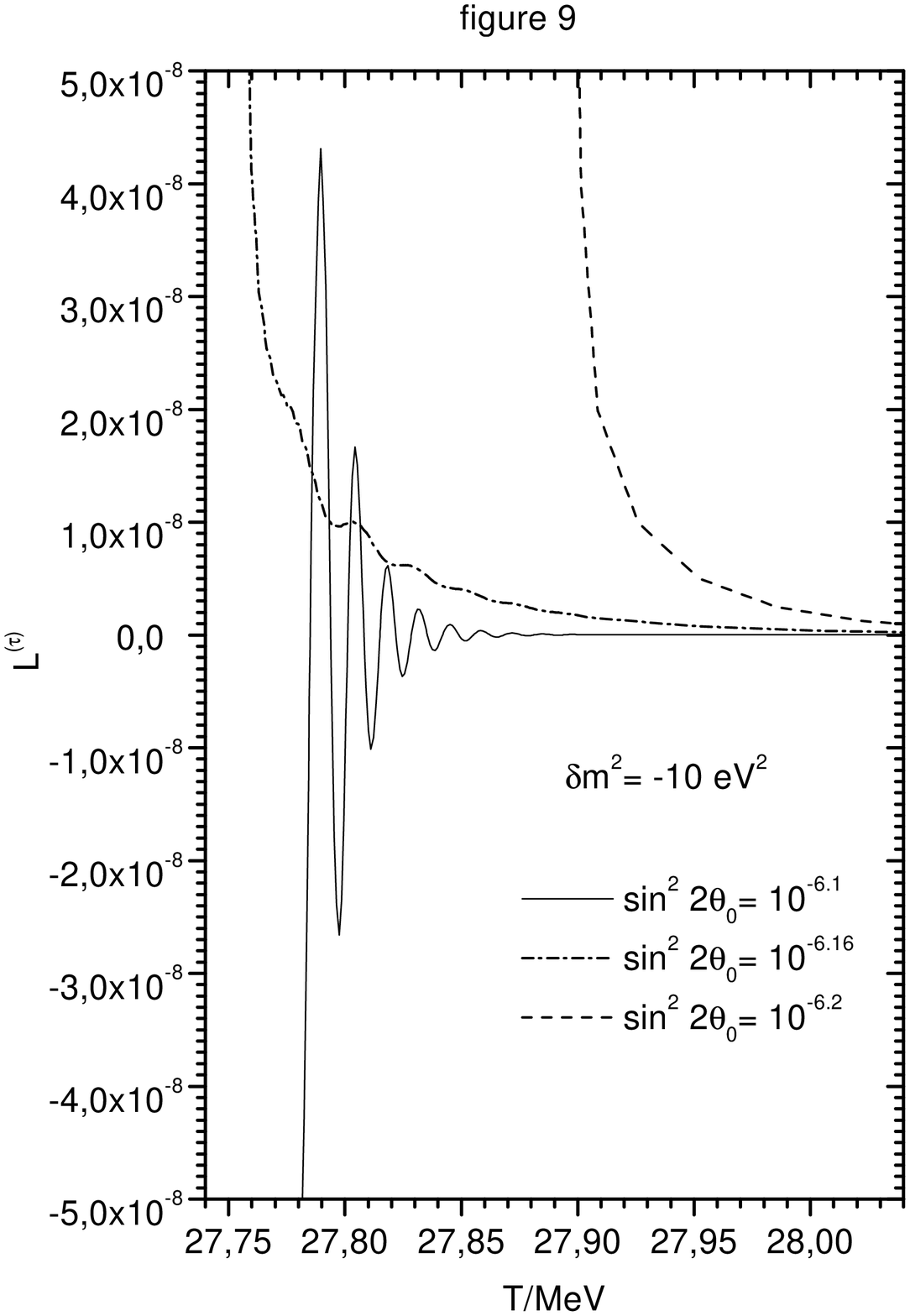,width=15cm}
\newpage
\epsfig{file=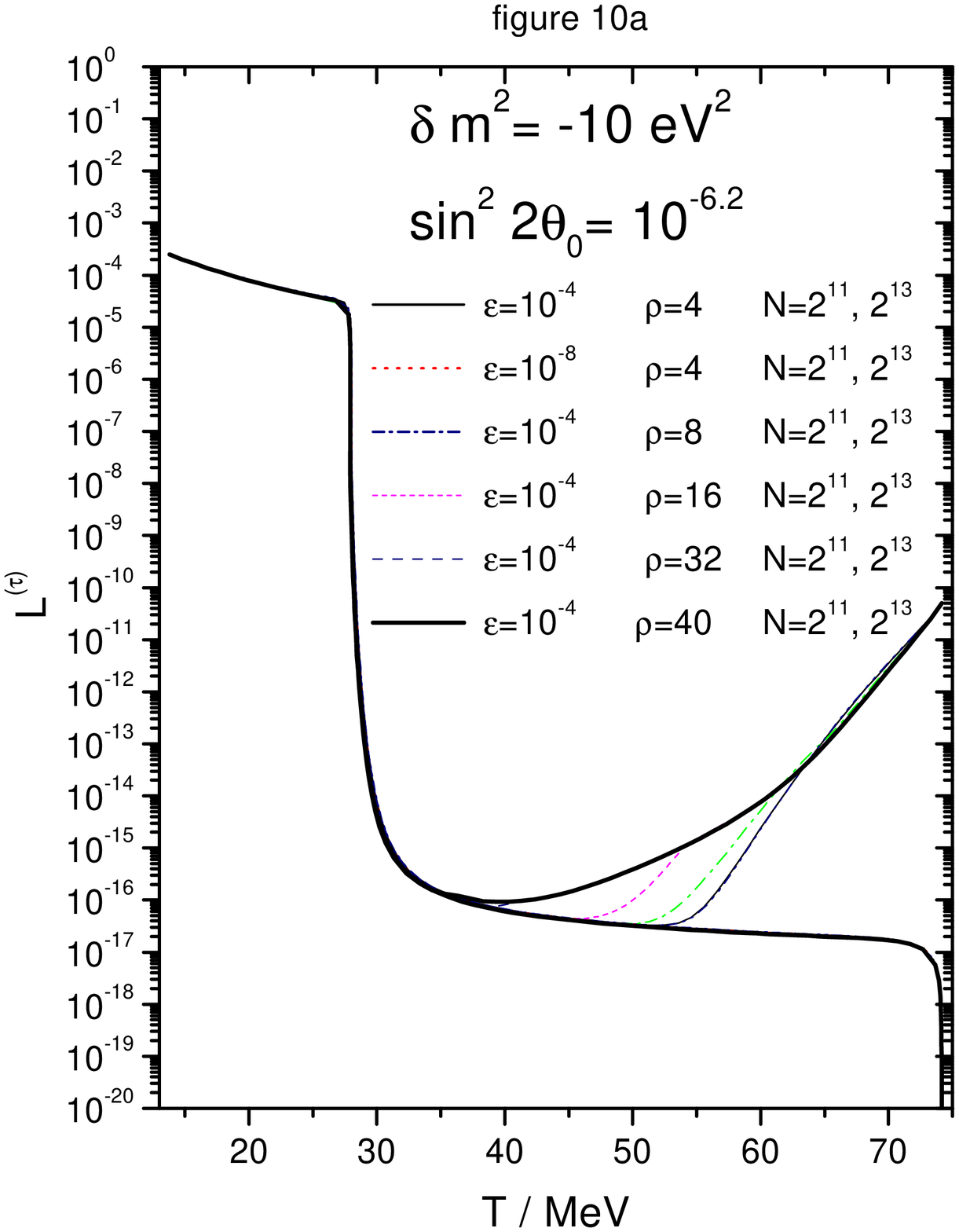,width=15cm}
\newpage
\epsfig{file=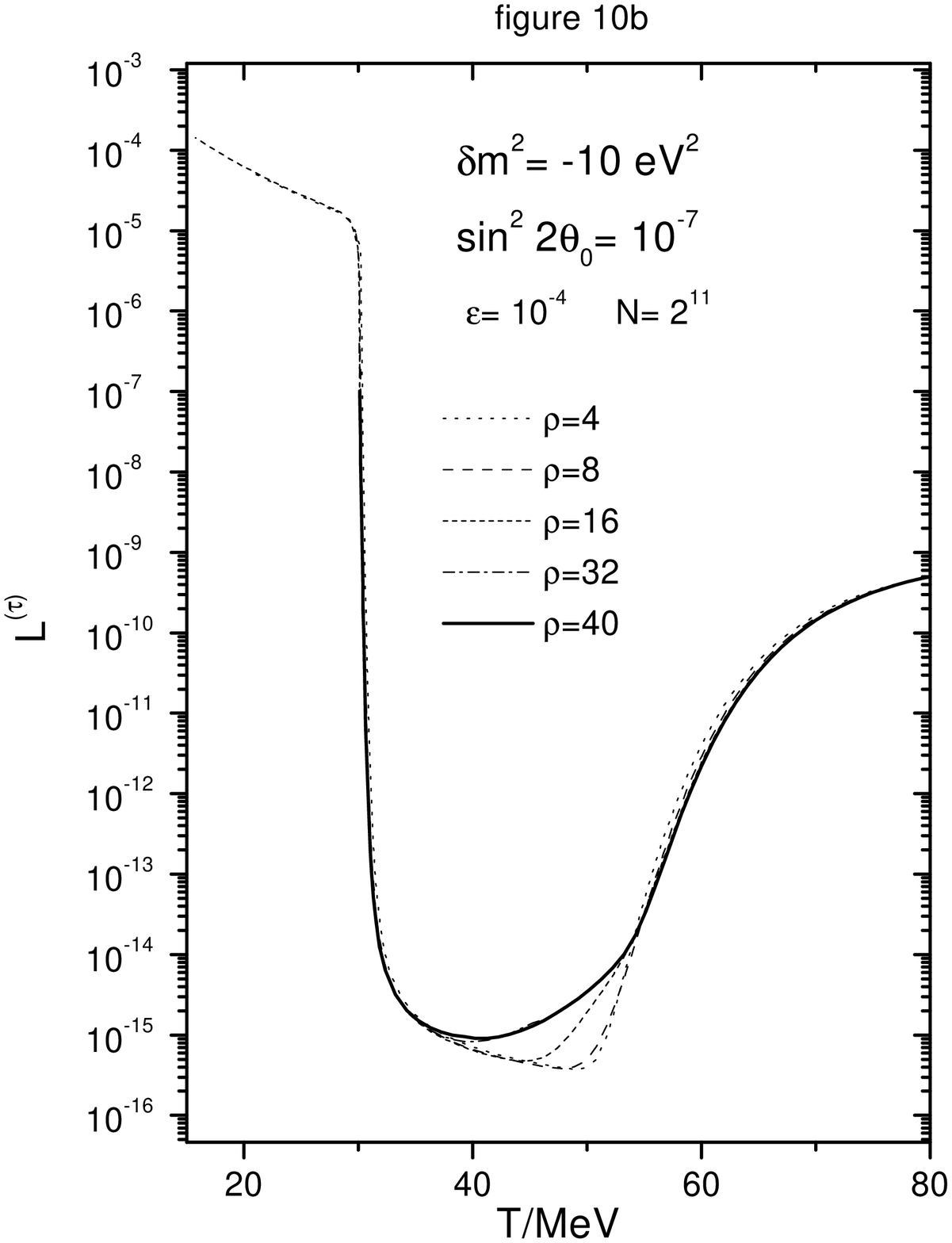,width=15cm}
\newpage
\epsfig{file=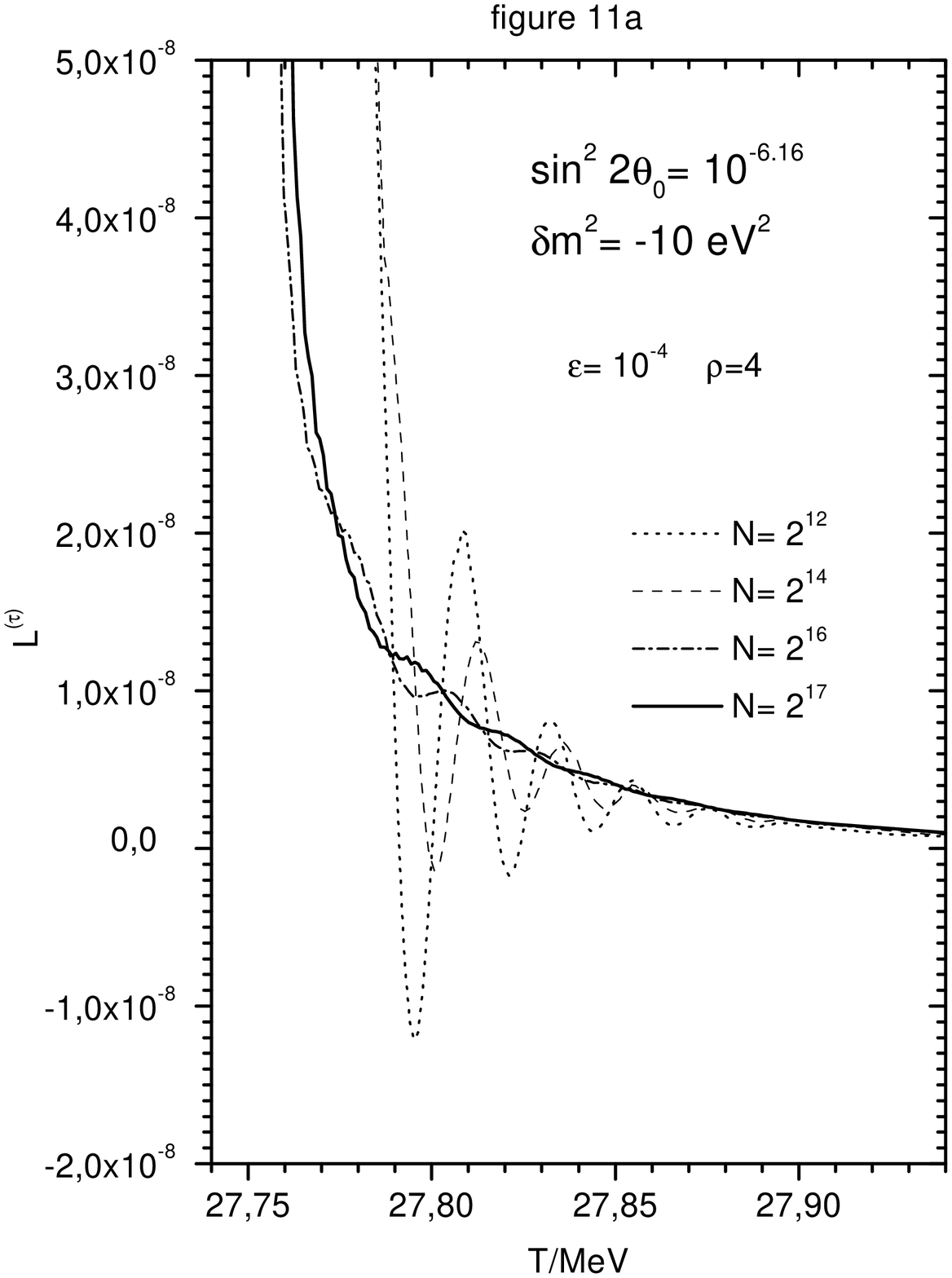,width=15cm}
\newpage
\epsfig{file=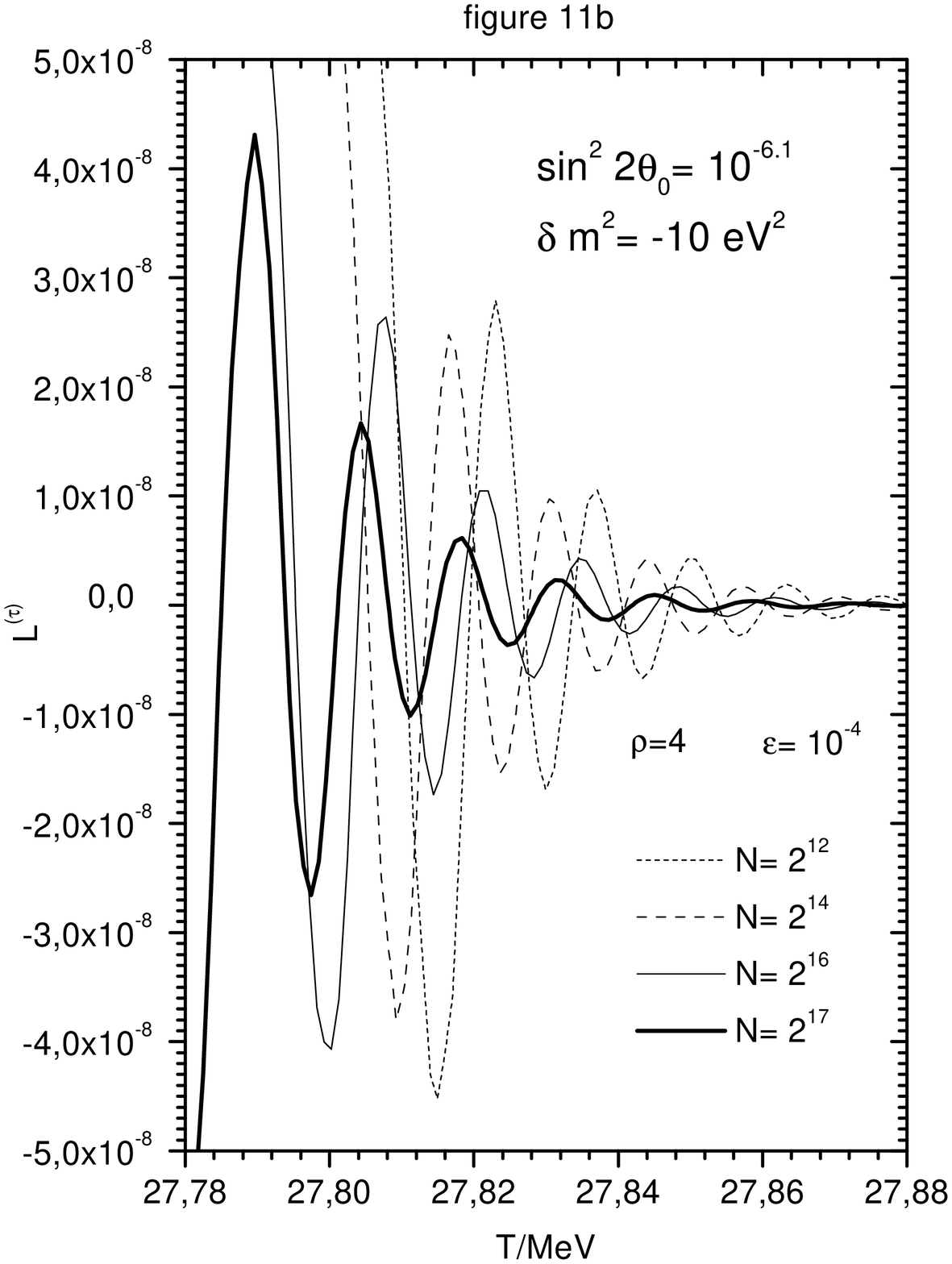,width=15cm}

\begin{thebibliography}{10}

\bibitem{ftv}
R. Foot, M. J. Thomson and R. R. Volkas, 
Phys. Rev. D53, 5349 (1996).

\bibitem{fv1}
R. Foot and R. R. Volkas, Phys. Rev. D55, 5147 (1997).

\bibitem{fv2} 
R. Foot and R. R. Volkas, Phys. Rev. D56, 6653 (1997);
Erratum-ibid, D59, 029901 (1999).

\bibitem{f}
R. Foot, Astropart. Phys. 10, 253 (1999).

\bibitem{bvw}
N. F. Bell, R. R. Volkas and Y.Y.Y.Wong, Phys. Rev. D59, 113001 (1999).

\bibitem{pas}
P. DiBari, P. Lipari and M. Lusignoli, hep-ph/9907548.

\bibitem{nfb}
For BBN implications of the neutrino asymmetry
in various 4 and 6 neutrino models,
see Ref.\cite{fv1,fv2,f,pas} and also
N. F. Bell, R. Foot and R. R. Volkas, Phys. Rev. D58, 105010 (1998);
R. Foot, hep-ph/9906311 (to appear in Phys. Rev. D). For
applications to models with mirror neutrinos, see
R. Foot and R. R. Volkas, Astropart. Phys. 7, 283 (1997);
hep-ph/9904336 (to appear in Phys. Rev. D).

\bibitem{barb}
K.Enqvist, K.Kainulainen and J.Maalampi,
\newblock {\em {N}ucl.{P}hys.} B349 754 (1991);
R.Barbieri and A.D. Dolgov,
\newblock {\em {N}ucl.{P}hys.} B349 743 (1991).

\bibitem{kir}
D.P.~Kirilova and M.V.~Chizhov, 
Nucl.Phys. B 534, 447 (1998).

\bibitem{dolgov}
A.D.Dolgov, S.H.Hansen, S.Pastor and D.V.Semikoz, hep-ph/9910444.

\bibitem{shifu}
X. Shi and G. Fuller, {\em Phys.Rev.Lett.} 83, 3120 (1999).

\bibitem{shi}
X. Shi, Phys. Rev. D54, 2753 (1996).

\bibitem{eks}
K. Enqvist, K. Kainulainen and A. Sorri, Phys. Lett.
B464, 199 (1999).

\bibitem{Sorri99}
A. Sorri, hep-ph/9911366.

\bibitem{stod}
R. A. Harris and L. Stodolsky, Phys. Lett. 116B, 464 (1982); 
Phys. Lett. B78, 313 (1978); 
A. Dolgov, Sov. J. Nucl. Phys. 33, 700 (1981).

\bibitem{bm} B. H. J. McKellar and M. J. Thomson, 
Phys. Rev. D49, 2710 (1994). 

\bibitem{stod2}
L. Stodolsky, Phys. Rev. D36, 2273 (1987);
M. Thomson, Phys. Rev. A45, 2243 (1991).

\bibitem{ekt}
K. Enqvist, K. Kainulainen and M. Thomson, Nucl. Phys. 
B 373, 498 (1992).

\bibitem{wolf78}
L. Wolfenstein, Phys. Rev. D17, 2369 (1978).

\bibitem{rn}
D. Notzold and G. Raffelt, Nucl. Phys. B307, 924 (1988).

\bibitem{ekm}
K. Enqvist, K. Kainulainen and J. Maalampi, Phys. Lett. 
B244, 186 (1990).

\bibitem{pas2}
P. Di Bari, hep-ph/9911214.

\bibitem{prl}
R. Foot and R. R. Volkas, Phys. Rev. Lett. 75, 4350 (1995).

\bibitem{lan}
G. N. Lance, `Numerical methods for high speed computers' (London 1960).

\bibitem{numrec}
W.H. Press, S.A.Teukolsky, W.T.Vetterling and B.P. Flannery,
`Numerical Recipes, the Art of Scientific Computing', 
Cambridge University Press (1994).


\end{thebibliography}
\end{document}